\documentstyle[aps,twocolumn,prb,epsf]{revtex} 

\begin{document}

\draft
\twocolumn[    
\hsize\textwidth\columnwidth\hsize\csname @twocolumnfalse\endcsname    

\title{Thermal transport in the Falicov-Kimball model
}
\author{J. K. Freericks$^{1}$ and V. Zlati\'c$^2$}
\address{$^1$Department of Physics, Georgetown University, 
  Washington, DC 20057-0995, U.S.A.\\
$^2$Institute of Physics, Zagreb, Croatia}
\date{\today}
\maketitle

\widetext
\begin{abstract}
We prove the Jonson-Mahan theorem for the thermopower of the Falicov-Kimball 
model by solving explicitly for the correlation functions in the large
dimensional limit. We prove a similar result for the thermal conductivity.
We separate the results
for thermal transport into the pieces of the heat current that arise from the
kinetic energy and those that arise from the potential energy.  
Our method of proof is specific to the Falicov-Kimball model, but 
illustrates the near cancellations between the kinetic-energy and 
potential-energy pieces of the heat current implied by the Jonson-Mahan theorem.
\end{abstract}

]      

\narrowtext
\section{Introduction}

The Jonson-Mahan\cite{jonson_mahan,mahan} theorem shows that there is 
a simple relation between the transport coefficient for the electrical
conductivity and that needed for the thermopower.  The relation is that
the integral for the $L_{12}$ coefficient has an extra power of
frequency in the integrand than the $L_{11}$ coefficient.  This result
has been known for many years\cite{mott} for a noninteracting system---the 
Jonson-Mahan
theorem generalizes this result for a wide class of many-body systems
(including the Falicov-Kimball model, the static Holstein model, and the
periodic Anderson model, but not including the Hubbard model or quantum Holstein
model).

We use the exact solution of the Falicov-Kimball model in the large-dimensional
limit to provide an alternate derivation of the Jonson-Mahan theorem by
explicitly evaluating all relevant correlation functions needed for 
the thermal transport.  Our exact analysis also allows us
to separate the contributions to thermal transport that arise 
from the kinetic energy and the potential energy pieces of the heat
current.  These results provide an interesting interpretation of
thermal transport in correlated systems.

In Section II we develop the formalism for deriving the dc conductivity, the
thermopower and the thermal conductivity.  We derive exact results for 
the relevant correlation functions and use them to prove the Jonson-Mahan
theorem and its generalization to the thermal
conductivity.  In Section III we provide numerical results for the thermal
transport illustrating the different contributions to the thermal coefficients
for a number of illustrative cases.  Conclusions are presented in Section IV.

\section{Formalism for the thermal transport}

The Hamiltonian for our system is the spin-one-half Falicov-Kimball 
model\cite{falicov_kimball}
\begin{equation}
H=-\frac{t^*}{2\sqrt{d}}\sum_{\langle i,j\rangle\sigma}c^\dagger_{i\sigma}\
c_{j\sigma}+E_f\sum_iw_i +U\sum_{i\sigma}w_ic^{\dagger}_{i\sigma}c_{i\sigma},
\label{eq: ham}
\end{equation}
where $c^{\dagger}_{i\sigma}$ ($c_{i\sigma})$ is the electron creation 
(annihilation)
operator for an electron at site $i$ with spin $\sigma$, $E_f$ is the energy
level of the localized electrons, $w_i$ is a variable that equals
zero or one and corresponds to the localized electron number, and $U$ is
the interaction strength.  The hopping integral is scaled with the spatial
dimension $d$ so as to have a finite result in the limit\cite{metzner_vollhardt}
$d\rightarrow \infty$; we measure all energies in units of $t^*=1$. We work
on a hypercubic lattice where the noninteracting density of states is
a Gaussian $\rho(\epsilon)=\exp(-\epsilon^2)/\sqrt{\pi}$.

The Falicov-Kimball model can be solved exactly by employing dynamical
mean field theory\cite{brandt_mielsch,freericks_fk}.  Because the self
energy $\Sigma(z)$ is local, the local Green's function satisfies
\begin{equation}
G(z)=\int d\epsilon \rho(\epsilon)\frac{1}{z+\mu-\Sigma(z)-\epsilon},
\label{eq: g_loc}
\end{equation}
with $z$ anywhere in the complex plane (we suppress the spin index here).  
The self energy, local Green's 
function, and effective medium $G_0$ are related by
\begin{equation}
G_0^{-1}(z)-G^{-1}(z)=\Sigma(z),
\label{eq: g_0}
\end{equation}
and the Green's function also satisfies
\begin{equation}
G(z)=(1-w_1)G_0(z)+w_1\frac{1}{G_0^{-1}(z)-U}.
\label{eq: g_atomic}
\end{equation}
Here $w_1$ is the average concentration of localized electrons,
\begin{equation}
w_1=2\exp[-\beta (E_f-\mu)]Z_\uparrow(\mu-U)Z_\downarrow(\mu-U)/Z,
\label{eq: w1}
\end{equation}
with $Z=Z_\uparrow(\mu)Z_\downarrow(\mu)+2\exp[-\beta(E_f-\mu)]Z_\uparrow(\mu-U)
Z_\downarrow(\mu-U)$ and
\begin{equation}
Z_\sigma(\mu)=2e^{\beta\mu/2}\prod_n \frac{i\omega_n+\mu-\lambda_\sigma
(i\omega_n)}{i\omega_n}.
\label{eq: z}
\end{equation}
The factor of two arises from the spin degeneracy of the f-electrons
and the constraint that
no more than one f-electron is allowed on any site.
The symbol $\lambda_\sigma(z)$ is defined from the effective medium via
$\lambda_\sigma(i\omega_n)=i\omega_n+\mu-G_{0\sigma}^{-1}(i\omega_n)$,
$\omega_n=\pi T(2n+1)$ is the Fermionic Matsubara frequency, and
$\beta=1/T$.  The algorithm for determining the Green's function begins with
the self energy set equal to zero.  Then Eq.~(\ref{eq: g_loc}) is used
to find the local Green's function.  The effective medium is found from
Eq.~(\ref{eq: g_0}) and the localized electron filling from Eq.~(\ref{eq: w1}).
The new local Green's function is then found from Eq.~(\ref{eq: g_atomic})
and the new self energy from Eq.~(\ref{eq: g_0}).  This algorithm is 
repeated until it converges.

Transport properties are calculated within a Kubo-Greenwood 
formalism\cite{kubo}.  This
relates the transport coefficients to correlation functions of the 
corresponding transport current operators.  We will deal with two
current operators here---the particle current\cite{mahan_text}
\begin{equation}
{\bf j}=\sum_{q\sigma}{\bf v}_q c^\dagger_{q\sigma}c_{q\sigma},
\label{eq: j_particle}
\end{equation}
(where the velocity operator is ${\bf v}_q=\nabla_q\epsilon(q)$ and the
Fourier transform of the creation operator is $c^\dagger_q=\sum_j
\exp[iq\cdot {\bf R}_j]c^\dagger_j$/N) and the heat 
current\cite{jonson_mahan,mahan_text}
\begin{eqnarray}
{\bf j}_Q&=&\sum_{q\sigma}(\epsilon_q-\mu){\bf v}_q c^\dagger_{q\sigma}c_{q\sigma}
\cr
&+&\frac{U}{2}\sum_{qq^\prime\sigma}W(q-q^\prime)[{\bf v}_q+{\bf v}_{q^\prime}]
c^\dagger_{q\sigma}c_{q^\prime\sigma},
\label{eq: j_heat}
\end{eqnarray}
[where  $W(q)=\sum_j\exp(-iq\cdot{\bf R}_j) w_j/N$]. The heat current can be
broken into two pieces: (i) a kinetic-energy piece ${\bf j}_Q^K$ which is
the first term in Eq.~(\ref{eq: j_heat}) and (ii) a potential energy piece
${\bf j}_Q^P$ which is the second term in Eq.~(\ref{eq: j_heat}). 

The particle current is defined by the commutator of the Hamiltonian
with the polarization operator.\cite{mahan_text}  When evaluated on a 
lattice with nearest-neighbor hopping, one finds factors that involve the
weighted summation of the nearest-neighbor translation vectors $\delta$
weighted by phase factors $\exp(iq\cdot\delta)$, which yield the velocity
operator terms above.  The definition of the heat-current operator is more
involved, and requires the Hamiltonian to be separated into operators $h_i$
that involve the site $i$ (in decomposing the kinetic-energy operator into
``localized'' pieces, one symmetrically assigns half of the $c_{i\sigma}^\dagger
c_{j\sigma}+c_{j\sigma}^\dagger c_{i\sigma}$ term
to site $i$ and half to site $j$).
These operators can be combined with the position operator to construct
an ``energy'' polarization operator $\sum_i {\bf R}_ih_i$, which is commuted
with the Hamiltonian to determine the energy current operator; the heat
current operator is just this energy current operator shifted by the
chemical potential multiplied by the number current operator.   
Important operator relations
between the heat-current operator and the particle-current operator are
described fully below [see the discussion around 
Eq.~(\ref{eq: heat_current_identity})].

The 
dc conductivity $\sigma$, thermopower $S$ and electronic thermal conductivity
$\kappa$ can all be determined from relevant correlation functions of the
current operators.  We define three transport coefficients $L_{11}$, $L_{12}=
L_{21}$, and $L_{22}$. Then
\begin{equation}
\sigma=\frac{e^2}{T}L_{11},
\label{eq: conductivity}
\end{equation}
\begin{equation}
S=-\frac{k_B}{|e|T}\frac{L_{12}}{L_{11}},
\label{eq: thermopower}
\end{equation}
and
\begin{equation}
\kappa=k_B^2\left [ L_{22}-\frac{L_{12}L_{21}}{L_{11}}\right ].
\label{eq: thermalconductivity}
\end{equation}
The transport coefficients are found from the analytic continuation of
the relevant ``polarization operators'' at zero frequency
\begin{eqnarray}
L_{11}&=&\lim_{\nu\rightarrow 0}{\rm Re}\frac{i}{\nu}\bar L_{11}(\nu),\cr
\bar L_{11}(i\nu_l)&=&\pi T\int_0^{\beta}d\tau e^{i\nu_l\tau}
\langle T_\tau j_{\alpha}^\dagger(\tau)j_{\beta}(0)\rangle,
\label{eq: l11}
\end{eqnarray}
where $\nu_l=2\pi Tl$ is the Bosonic Matsubara frequency, the $\tau$-dependence
of the operator is with respect to the full Hamiltonian in Eq.~(\ref{eq: ham}),
and we must analytically continue $\bar L_{11}(i\nu_l)$ to the real
axis $\bar L_{11}(\nu)$ before taking the limit $\nu\rightarrow 0$.  Similar
definitions hold for the other transport coefficients:
\begin{eqnarray}
L_{12}&=&\lim_{\nu\rightarrow 0}{\rm Re}\frac{i}{\nu}\bar L_{12}(\nu),\cr
\bar L_{12}(i\nu_l)&=&\pi T\int_0^{\beta}d\tau e^{i\nu_l\tau}
\langle T_\tau j_{\alpha}^\dagger(\tau)j_{Q\beta}(0)\rangle,
\label{eq: l12}
\end{eqnarray}
and
\begin{eqnarray}
L_{22}&=&\lim_{\nu\rightarrow 0}{\rm Re}\frac{i}{\nu}\bar L_{22}(\nu),\cr
\bar L_{22}(i\nu_l)&=&\pi T\int_0^{\beta}d\tau e^{i\nu_l\tau}
\langle T_\tau j_{Q\alpha}^\dagger(\tau)j_{Q\beta}(0)\rangle.
\label{eq: l22}
\end{eqnarray}
In all of these equations, the subscripts $\alpha$ and $\beta$ denote the
respective spatial index of the current vectors.

We begin with a derivation that shows the analytic continuation for the
conductivity.  Substituting the definition of the particle current
operator of Eq.~(\ref{eq: j_particle}) into Eq.~(\ref{eq: l11}) for
$\bar L_{11}$ yields
\begin{eqnarray}
\bar L_{11}(i\nu_l)&=&\pi 
T\int_0^\beta d\tau e^{i\nu_l\tau}\sum_{qq^\prime\sigma
\sigma^\prime}{\bf v}_{q\alpha}{\bf v}_{q^\prime\beta}\cr
&\times&\langle T_{\tau}
c^\dagger_{q\sigma}(\tau)c_{q\sigma}(\tau)c^\dagger_{q^\prime\sigma^\prime}(0)
c_{q^\prime\sigma^\prime}(0)\rangle.
\label{eq: l11_v}
\end{eqnarray}
The correlation function can be determined from Dyson's equation which
relates the dressed correlation function to the bare correlation function
via the irreducible charge vertex.  Since the charge vertex is local in
the infinite-dimensional limit, it is an even function of momentum, and
any sum over momentum that is weighted by just one factor of ${\bf v}_q$
will vanish.  Hence, the dressed correlation function is equal to just the
bare correlation function\cite{khurana} (note that the contractions of the 
operators at equal times also vanish when summed over momentum).  This produces
\begin{equation}
\bar L_{11}(i\nu_l)=-\pi T\int_0^\beta d\tau e^{i\nu_l\tau}\sum_{q\sigma}
{\bf v}_{q\alpha}{\bf v}_{q\beta}G_{q\sigma}(\tau)G_{q\sigma}(-\tau).
\label{eq: l11_bare}
\end{equation}
Now, we introduce the Fourier transform of the Green's function
$G(\tau)=T\sum_n\exp(-i\omega_n\tau)G_n$ with $G_n=\int_0^\beta d\tau 
\exp(i\omega_n\tau )G(\tau)$.  Substituting into Eq.~(\ref{eq: l11_bare}),
allows us to perform the integral over $\tau$.  This finally produces
\begin{equation}
\bar L_{11}(i\nu_l)=-\pi 
T^2\sum_n\sum_{q\sigma}{\bf v}_{q\alpha}{\bf v}_{q\beta}
G_{q\sigma}(i\omega_n)G_{q\sigma}(i\omega_{n+l}).
\label{eq: l11_imag}
\end{equation}

\begin{figure}[htbf]
\epsfxsize=2.5in
\centerline{\epsffile{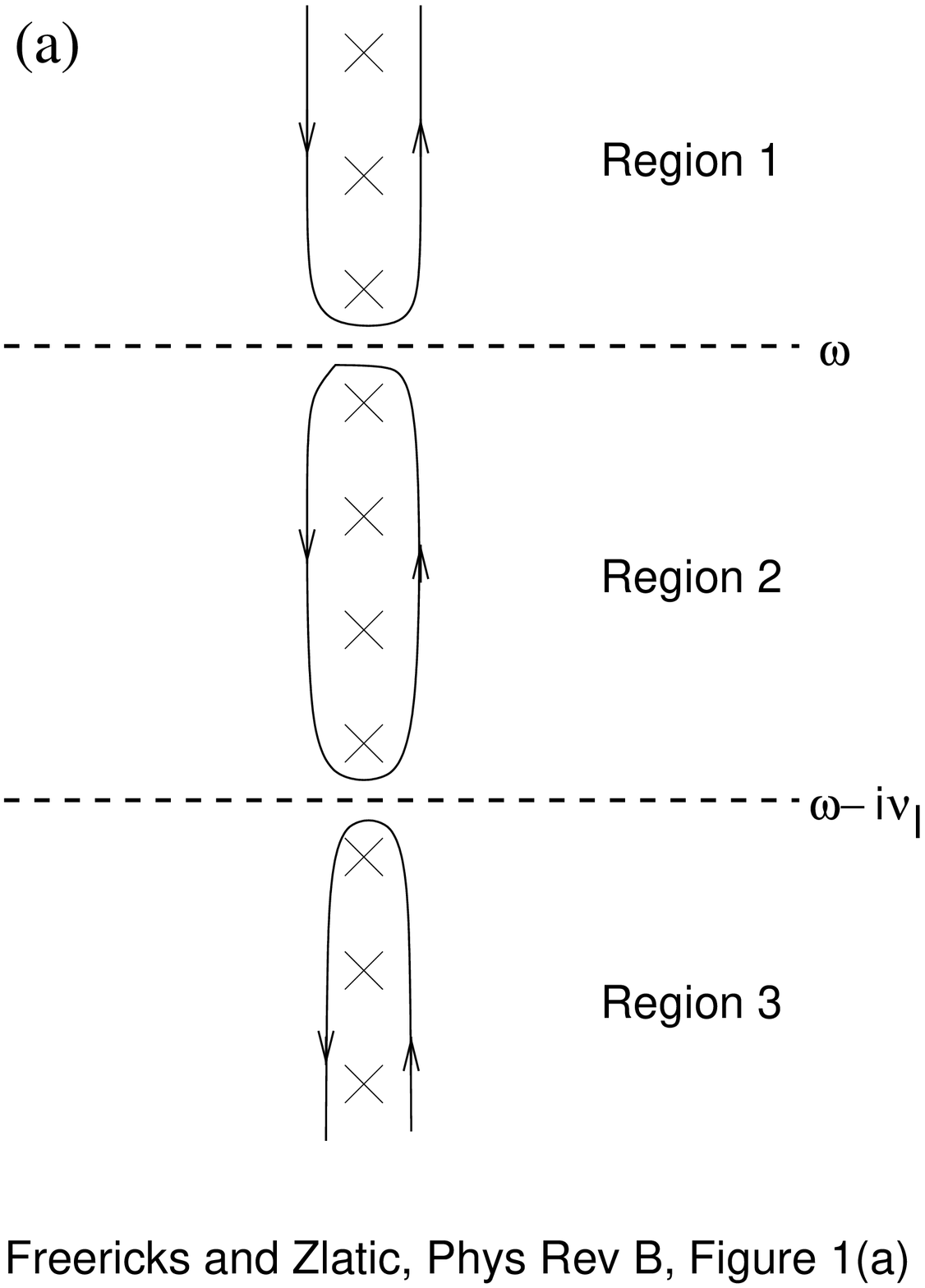}}
\epsfxsize=2.5in
\centerline{\epsffile{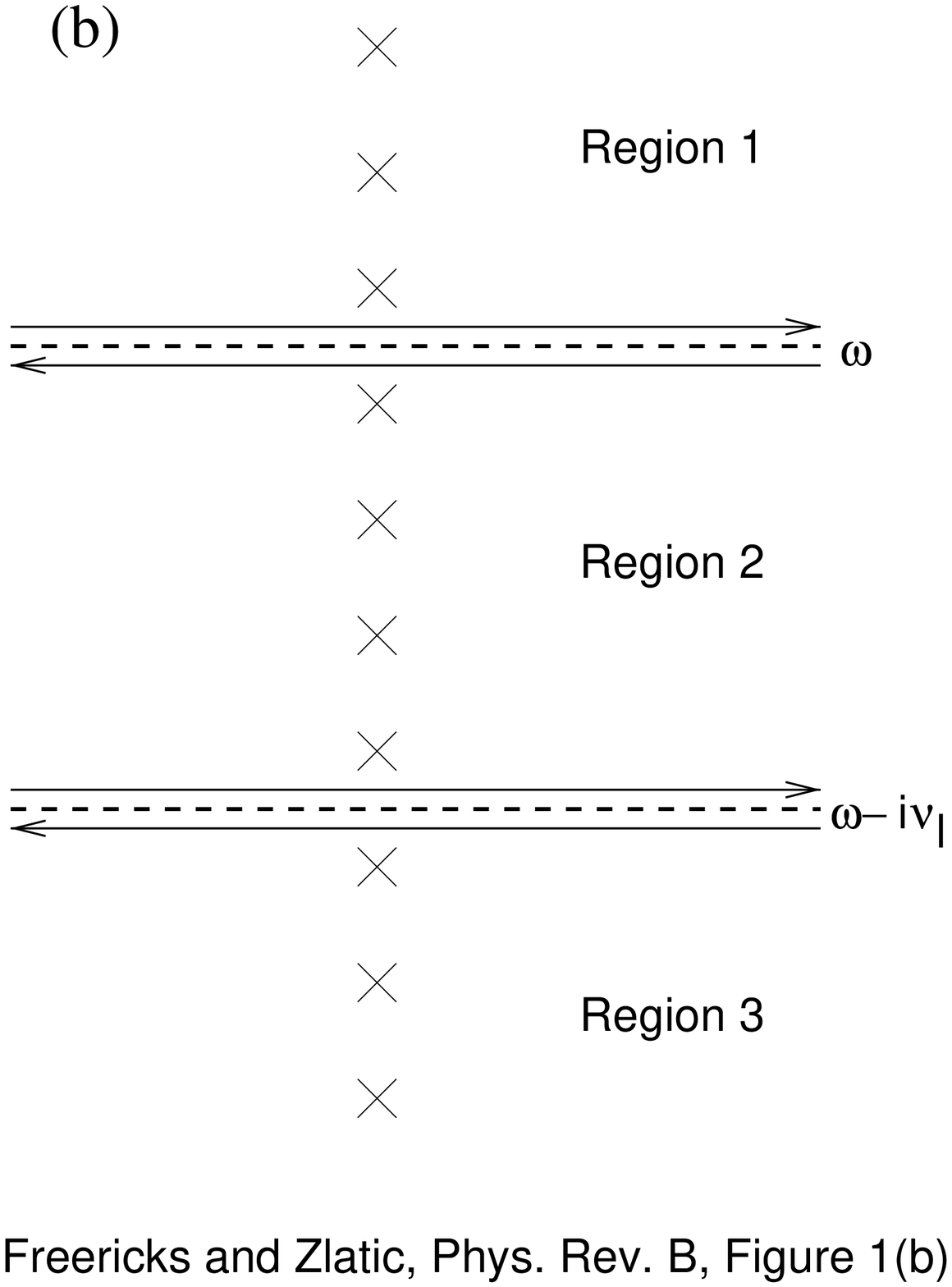}}
\caption{ Contours used in the analytic continuation: (a) contour needed
for the Matsubara frequency summation and (b) deformed contour to lines
parallel to the real axis.}
\label{fig: contours}
\end{figure}

The next step is to perform the analytic continuation from the imaginary to
real axis.  The procedure is standard\cite{mahan_text}.  We first write
the summation over Matsubara frequencies as an integral over the contour
$C$ shown in Fig.~1(a) which has contributions at the poles of the Fermi
function $f(\omega)=1/[1+\exp(\beta\omega)]$ which lie at the Fermionic 
Matsubara frequencies.
The contours are then deformed to lines parallel to the real axis, with
the Green's functions evaluated with either retarded (R) or advanced (A)
functions. The result is
\begin{eqnarray}
\bar L_{11}(i\nu_l)&=&-\frac{T}{2i}\int_C d\omega f(\omega)
\sum_{q\sigma}{\bf v}_{q\alpha}{\bf v}_{q\beta}\cr
&~&\times G_{q\sigma}(\omega)G_{q\sigma}(\omega+i\nu_l),\cr
&=&-\frac{T}{2i}\int_{-\infty}^{\infty}d\omega f(\omega)
\sum_{q\sigma}{\bf v}_{q\alpha}{\bf v}_{q\beta}\cr
&~&\times
[G^R_{q\sigma}(\omega)-G^A_{q\sigma}(\omega)]G^R_{q\sigma}(\omega+i\nu_l) \cr
&-&\frac{T}{2i}\int_{-\infty}^{\infty}d\omega f(\omega-i\nu_l)
\sum_{q\sigma}{\bf v}_{q\alpha}{\bf v}_{q\beta}\cr
&~&\times
G^A_{q\sigma}(\omega-i\nu_l)[G^R_{q\sigma}(\omega)-G^A_{q\sigma}(\omega)].
\label{eq: l11_contour}
\end{eqnarray}
The analytic continuation is performed by first rewriting $ f(\omega-i\nu_l)=
 f(\omega)$, then taking $i\nu_l\rightarrow \nu+i\delta$ and shifting the
integration variable $\omega\rightarrow\omega+\nu$ in the second
integral.  Then, using the definition for $L_{11}$, we finally arrive at
\begin{eqnarray}
L_{11}&=&\lim_{\nu\rightarrow 0}-\frac{T}{2\nu}\int_{-\infty}^{\infty}d\omega
\sum_{q\sigma}{\bf v}_{q\alpha}{\bf v}_{q\beta}\cr
&\times&{\rm Re}\biggr\{ f(\omega)G_{q\sigma}(\omega)G_{q\sigma}(\omega+\nu)\cr
&-&f(\omega+\nu)G^*_{q\sigma}(\omega)G^*_{q\sigma}(\omega+\nu)\cr
&-&[f(\omega)-f(\omega+\nu)]G^*_{q\sigma}(\omega)G_{q\sigma}(\omega+\nu)
\biggr\}.
\label{eq: l11_real}
\end{eqnarray}
Since $G_{q\sigma}(\omega)=1/[\omega+\mu-\Sigma(\omega)-\epsilon(q)]$, 
we can perform the summation over $q$ directly.  Because $\epsilon(q)$
is an even function of $q$ and ${\bf v}_q$ is odd, we must have $\alpha=\beta$.
Converting the fraction into the integral of an exponential, then allows
the summation over $q$ to be performed directly\cite{muller-hartmann}. The
summation over $q$ can be written as an integral over energy with a
weighting factor of $\rho(\epsilon)t^{*2}/d$.  This yields
\begin{eqnarray}
L_{11}&=&\lim_{\nu\rightarrow 0}-\frac{Tt^{*2}}{2\nu d}\delta_{\alpha\beta}
\int_{-\infty}^{\infty}d\omega \sum_{\sigma}\int_{-\infty}^{\infty}d\epsilon
\rho(\epsilon)\cr
&\times&{\rm Re}\biggr\{ f(\omega)G_{q\sigma}(\omega)G_{q\sigma}(\omega+\nu)\cr
&-&f(\omega+\nu)G^*_{q\sigma}(\omega)G^*_{q\sigma}(\omega+\nu)\cr
&-&[f(\omega)-f(\omega+\nu)]G^*_{q\sigma}(\omega)G_{q\sigma}(\omega+\nu)
\biggr\}.
\label{eq: l11_real2}
\end{eqnarray}
If we define $\sigma_0=e^2t^{*2}/(2d)$, and we perform the integral 
over $\epsilon$, we arrive at 
\begin{eqnarray}
L_{11}&=&\lim_{\nu\rightarrow 0}\frac{T\sigma_0}{e^2}\delta_{\alpha\beta} 
\int_{-\infty}^{\infty}d\omega \cr
&\times&\frac{f(\omega)-f(\omega+\nu)}{\nu}{\rm Re}\biggr [
-\frac{G(\omega)-G(\omega+\nu)}{\nu+\Sigma(\omega)-\Sigma(\omega+\nu)}
\cr
&+&\frac{G^*(\omega)-G(\omega+\nu)}{\nu+\Sigma^*(\omega)-\Sigma(\omega+\nu)}
\biggr ].
\label{eq: l11_real3}
\end{eqnarray}
The final step is to take the limit of $\nu\rightarrow 0$. Using the facts that
\begin{equation}
\lim_{\nu\rightarrow 0}\frac{f(\omega)-f(\omega+\nu)}{\nu}=
-\frac{df(\omega)}{d\omega},
\label{eq: f_deriv}
\end{equation}
and
\begin{equation}
\lim_{\nu\rightarrow 0}\frac{G(\omega)-G(\omega+\nu)}
{\nu+\Sigma(\omega)-\Sigma(\omega+\nu)}=-2+2[\omega+\mu-\Sigma(\omega)]
G(\omega),
\label{eq: g_deriv}
\end{equation}
produces our final result for $L_{11}$
\begin{equation}
L_{11}=\frac{T\sigma_0}{e^2}\int_{-\infty}^{\infty}d\omega
\left ( -\frac{df(\omega)}{d\omega} \right ) \tau(\omega),
\label{eq: l11_final}
\end{equation}
with the relaxation time $\tau(\omega)$ defined by
\begin{equation}
\tau(\omega)=\frac{{\rm Im}G(\omega)}{{\rm Im}\Sigma(\omega)}+2-2
{\rm Re}\{ [\omega+\mu-\Sigma(\omega)]G(\omega)\}.
\label{eq: tau}
\end{equation}
This result appears different from that originally derived for the
hypercubic lattice\cite{jarrell_cond}, but Eqs.~(\ref{eq: l11_final})
and ({\ref{eq: tau}) do yield the same result; the form presented here has
the integral over the noninteracting density of states performed exactly.

The derivation for the transport coefficient $L_{12}$, needed for the
thermopower, proceeds in a similar fashion.  We can divide the heat current
into two pieces, one corresponding to the kinetic energy and one corresponding
to the potential energy.  This allows us to write $L_{12}=L_{12}^K+L_{12}^P$.
The derivation for the kinetic energy piece follows exactly like the 
derivation for 
$L_{11}$ except there is an extra factor of $\epsilon-\mu$ that appears
in Eq.~(\ref{eq: l11_real2}).  The integral over $\epsilon$
can then be performed straightforwardly producing
\begin{eqnarray}
&~&L^K_{12}=\lim_{\nu\rightarrow 0}\frac{T\sigma_0}{e^2}\delta_{\alpha\beta}
\int_{-\infty}^{\infty}d\omega
\frac{f(\omega)-f(\omega+\nu)}{\nu}\cr
&\times&{\rm Re}\biggr \{
-\frac{[\omega-\Sigma(\omega)]G(\omega)-[\omega+\nu-\Sigma(\omega+\nu)]
G(\omega+\nu)}{\nu+\Sigma(\omega)-\Sigma(\omega+\nu)}
\cr
&+&\frac{[\omega-\Sigma^*(\omega)]G^*(\omega)-[\omega+\nu-\Sigma(\omega+\nu)]
G(\omega+\nu)}{\nu+\Sigma^*(\omega)-\Sigma(\omega+\nu)}
\biggr \}.
\label{eq: l12k_real}
\end{eqnarray} 
Evaluating the limit $\nu\rightarrow 0$ is simple.  The final result is
\begin{eqnarray}
L^K_{12}&=&\frac{T\sigma_0}{e^2}\int_{-\infty}^{\infty}d\omega
\left ( -\frac{df(\omega)}{d\omega} \right ) 
\{ [\omega-{\rm Re}\Sigma(\omega)]
\tau(\omega) \cr
&-&2{\rm Im}\Sigma(\omega){\rm Im}[(\omega+\mu-\Sigma(\omega))
G(\omega)]\},
\label{eq: l12k_final}
\end{eqnarray}
with $\tau(\omega)$ defined in Eq.~(\ref{eq: tau}).

The derivation of the potential energy piece is much more involved.
The first step is to replace the momentum-dependent operator
$W(q-q^\prime)$ by its Fourier transform.  Simplifying the expression for
$\bar L_{12}^P$ yields
\begin{eqnarray}
\bar L_{12}^P(i\nu_l)&=&\frac{\pi TU}{2}\int_0^\beta d\tau e^{i\nu_l\tau}
\sum_{qq^\prime\sigma\sigma^\prime}\frac{1}{N}
\sum_j{\bf v}_{q\alpha}{\bf v}_{q^\prime \beta} \cr
&\times&\biggr [ e^{-iq^\prime\cdot {\bf R}_j}
\langle T_\tau
w_j c^\dagger_{q\sigma}(\tau)c_{q\sigma}(\tau)c^{\dagger}_{q^\prime
\sigma^\prime}(0)c_{j\sigma^\prime}(0)\rangle\cr
&+&e^{iq^\prime\cdot {\bf R}_j}
\langle T_\tau w_j c^\dagger_{q\sigma}(\tau)c_{q\sigma}(\tau)c^{\dagger}_{j
\sigma^\prime}(0)c_{q^\prime\sigma^\prime}(0)\rangle \biggr ].
\label{eq: l12p}
\end{eqnarray}
Noting that the $w_j$ operator commutes with the Fermionic operators,
allows us to use Wick's theorem to rewrite
the terms in the square bracket as  
\begin{eqnarray}
\frac{1}{N}\sum_{j}&\biggr [& -e^{-iq\cdot {\bf R}_j}
\langle T_\tau
w_j c^\dagger_{q\sigma}(\tau)c_{j\sigma}(0)\rangle G_{q\sigma}(\tau)\cr
&+&e^{iq\cdot {\bf R}_j}
\langle T_\tau w_j c_{q\sigma}(\tau)c^{\dagger}_{j
\sigma}(0)\rangle  G_{q\sigma}(-\tau)\biggr ]\delta_{qq^\prime}\delta_{\sigma
\sigma^\prime},
\label{eq: l12p_wick}
\end{eqnarray}
where we have taken the appropriate contractions (note the velocity operators
guarantee that we need not worry about any vertex corrections).
The correlation functions in Eq.~(\ref{eq: l12p_wick}) can be evaluated by
taking the derivative with respect to the components of an
infinitesimal field $-\sum_j h_j w_j$.  These
correlation functions have a factor of $\exp[-\beta H]$ in the numerator
and a factor of $Z$ in the denominator.  In addition, the $\tau$-dependence
of the operators arises from factors of $\exp [\pm \tau H]$.  Since the
operator $w_j$ commutes with all Fermionic operators, it is easy to
verify that the expression in Eq.~(\ref{eq: l12p_wick}) becomes
\begin{eqnarray}
-\frac{1}{N}\sum_j &\biggr [&G_{q\sigma}(\tau)
\left ( T\frac{\partial}{\partial h_j}+\langle w_j
\rangle \right ) G_{q\sigma}(-\tau)\cr
&+&  G_{q\sigma}(-\tau) \left ( T\frac{\partial}{\partial h_j}+\langle w_j
\rangle \right ) G_{q\sigma}(\tau)\biggr ] 
\delta_{qq^\prime}\delta_{\sigma   \sigma^\prime},
\label{eq: correlation}
\end{eqnarray}
which follows by first removing the $w_j$ operator through the derivative, then
expressing the Fermionic operator at site $j$ through a Fourier transform, 
and finally evaluating the Fermionic averages.
Substituting this
result into Eq.~(\ref{eq: l12p}) then yields 
\begin{eqnarray}
\bar L_{12}^P(i\nu_l)&=&-\frac{\pi T^2U}{2N}
\sum_n\sum_{q\sigma}\sum_j{\bf v}_{q\alpha}
{\bf v}_{q\beta} \cr
&\times&\Biggr ( \{ [T\frac{\partial}{\partial h_j}+\langle w_j\rangle ]
G_{q\sigma}(i\omega_n)\}G_{q\sigma}(i\omega_{n+l})\cr
&+&G_{q\sigma}(i\omega_n)[T\frac{\partial}{\partial h_j}+\langle w_j\rangle ]
G_{q\sigma}(i\omega_{n+l})\Biggr ) .
\label{eq: l12p_imag}
\end{eqnarray}
The derivatives need to be computed. Writing the momentum-dependent Green's
function as a Fourier transform 
\begin{equation}
G_{q\sigma}(i\omega_n)=\frac{1}{N}\sum_{i-j}e^{iq\cdot({\bf R}_i-{\bf R}_j)}
G_{ij\sigma}(i\omega_n),
\label{eq: g_fourier}
\end{equation}
and using the identity 
\begin{equation}
G_{ij\sigma}(i\omega_n)=\sum_{kl}G_{ik\sigma}(i\omega_n)G_{kl\sigma}^{-1}
(i\omega_n)G_{lj\sigma}(i\omega_n),
\label{eq: g_identity}
\end{equation}
(with $G^{-1}$ the matrix inverse of $G$) 
allows us to compute the derivative as
\begin{eqnarray}
\frac{\partial}{\partial h_j}G_{q\sigma}(i\omega_n)&=&\frac{1}{N}\sum_{i-j}
e^{iq\cdot({\bf R}_i-{\bf R}_j)} G_{ij\sigma}(i\omega_n)\cr
&\times& \frac{\partial \Sigma_{jj\sigma}(i\omega_n)}{\partial h_j}
G_{jj\sigma}(i\omega_n).
\label{eq: g_deriv_sum}
\end{eqnarray}
But in a homogeneous phase, the derivative of the local self energy with
respect to the local field, and the local Green's function are both
independent of the site $j$, so we finally arrive at
\begin{equation}
\frac{\partial G_{q\sigma}(i\omega_n)}{\partial h}=G_{n\sigma}\frac{\partial 
\Sigma_{n\sigma}}
{\partial h}G_{q\sigma}(i\omega_n).
\label{eq: sigma_deriv}
\end{equation}
Since the self energy depends only on $G_n$ and $w_1$, the derivative
can be computed by taking partial derivatives and using the chain rule
\begin{equation}
\frac{\partial \Sigma_{n\sigma}}{\partial h}=
\frac{\frac{\partial \Sigma_{n\sigma}}{\partial w_1}\frac{\partial w_1}
{\partial h}}{1-G^2_{n\sigma}\frac{\partial \Sigma_{n\sigma}}{\partial
G_{n\sigma}}}.
\label{eq: sigma_deriv2}
\end{equation}
Each of the derivatives in Eq.~(\ref{eq: sigma_deriv2}) can be found
directly\cite{brandt_mielsch,freericks_fk}
\begin{equation}
\frac{\partial \Sigma_{n\sigma}}{\partial w_1}=\frac{U}{1+G_{n\sigma}
(2\Sigma_{n\sigma}-U)},
\label{eq: sigma_deriv_w1}
\end{equation}
\begin{equation}
\frac{\partial w_1}{\partial h}=\frac{w_1(1-w_1)}{T},
\label{eq: w1_deriv}
\end{equation}
and
\begin{equation}
1-G^2_{n\sigma}\frac{\partial \Sigma_{n\sigma}}{\partial G_{n\sigma}}
=\frac{(1+G_{n\sigma}\Sigma_{n\sigma})(1+G_{n\sigma}[\Sigma_{n\sigma}-U])}
{1+G_{n\sigma}(2\Sigma_{n\sigma}-U)}.
\label{eq: sigma_deriv_g}
\end{equation}
Substituting these derivatives into Eq.~(\ref{eq: sigma_deriv2}) and performing
some straightforward simplifications that involve the quadratic equation
that the self energy satisfies\cite{brandt_mielsch} finally yields
\begin{equation}
[T\frac{\partial}{\partial h}+\langle w\rangle]G_{q\sigma}(i\omega_n)=
\frac{\Sigma_{n\sigma}}{U}G_{q\sigma}(i\omega_n).
\label{eq: field_deriv}
\end{equation}

Now we are ready to perform the analytic continuation.  First we substitute
Eq.~(\ref{eq: field_deriv}) into Eq.~(\ref{eq: l12p_imag}) and we note that
the sum over $j$ cancels the factor of $1/N$
\begin{eqnarray}
\bar L^P_{12}(i\nu_l)&=&-\frac{\pi
T^2}{2}\sum_n\sum_{q\sigma}{\bf v}_{q\alpha}{\bf v}_{q\beta}
\cr
&\times&[\Sigma_\sigma(i\omega_n)+\Sigma_\sigma(i\omega_{n+l})]
G_{q\sigma}(i\omega_n)G_{q\sigma}(i\omega_{n+l}).
\label{eq: l12p_imag2}
\end{eqnarray}
Next, we rewrite the sum over Matsubara frequencies as a contour integral
and perform the analytic continuation in the exact same way as before.  If
we then evaluate $L_{12}^P$ we find
\begin{eqnarray}
L^P_{12}&=&\lim_{\nu\rightarrow 0}-\frac{T\sigma_0}{2e^2\nu}
\int_{-\infty}^\infty d\omega \int_{-\infty}^\infty d\epsilon \rho(\epsilon)\cr
&\times&{\rm Re}\biggr [ f(\omega)\{ [\Sigma(\omega)+\Sigma(\omega+\nu)]
G_q(\omega)G_q(\omega+\nu)\cr
&-&[\Sigma^*(\omega)+\Sigma(\omega+\nu)]
G^*_q(\omega)G_q(\omega+\nu)\}\cr
&+&f(\omega+\nu)\{ [\Sigma^*(\omega)+\Sigma(\omega+\nu)]
G^*_q(\omega)G_q(\omega+\nu)\cr
&-&[\Sigma^*(\omega)+\Sigma^*(\omega+\nu)]
G^*_q(\omega)G^*_q(\omega+\nu)\} \biggr ].
\label{eq: l12p_real}
\end{eqnarray}
Now the integral over $\epsilon$ can be performed and the limit 
$\nu\rightarrow 0$ can be taken.  It becomes
\begin{eqnarray}
L^P_{12}&=&\frac{T\sigma_0}{e^2}\int_{-\infty}^\infty d\omega
\left ( -\frac{d f(\omega)}{d\omega}\right ) \biggr [ {\rm Re}\Sigma(\omega)
\tau(\omega)\cr
&+&2{\rm Im}\Sigma(\omega){\rm Im}[\{\omega+\mu-\Sigma(\omega)\}G(\omega)]
\biggr ].
\label{eq: l12p_final}
\end{eqnarray}
Adding together Eqs.~(\ref{eq: l12k_final}) and (\ref{eq: l12p_final}) 
yields the Jonson-Mahan result of
\begin{equation}
L_{12}=\frac{T\sigma_0}{e^2}\int_{-\infty}^\infty d\omega
\left ( -\frac{d f(\omega)}{d\omega}\right ) \tau(\omega)\omega .
\label{eq: l12_final}
\end{equation}

Our final derivation is for the thermal conductivity coefficient $L_{22}$.  Like
before, we separate this into pieces corresponding to the kinetic energy
and the potential energy: $L_{22}=L_{22}^{KK}+L_{22}^{KP}+L_{22}^{PK}+
L_{22}^{PP}$.  Due to the symmetry of the terms, we have $L_{22}^{KP}=
L_{22}^{PK}$.  The kinetic energy piece is simple to calculate.  Like in
our derivation for $L_{12}^K$, the steps are identical to the derivation
for $L_{11}$ except we have an extra factor of $(\epsilon-\mu)^2$ in
Eq.~(\ref{eq: l11_real2}).  Performing the integration over $\epsilon$
and collecting terms finally yields
\begin{eqnarray}
L_{22}^{KK}&=&\frac{T\sigma_0}{e^2}\int_{-\infty}^{\infty}d\omega
\left ( -\frac{df(\omega)}{d\omega}\right )\cr
&\times&\biggr \{ [\omega-{\rm Re}\Sigma(\omega)]^2\tau(\omega)+{\rm Im}
G(\omega){\rm Im}\Sigma(\omega)\cr
&-&2[{\rm Im}\Sigma(\omega)]^2+2[{\rm Im}\Sigma(\omega)]^2{\rm Re} \{
[\omega+\mu-\Sigma(\omega)]G(\omega)\}\cr
&-&4[\omega-{\rm Re}\Sigma(\omega)]{\rm Im}\Sigma(\omega){\rm Im}
\{[\omega+\mu-\Sigma(\omega)]G(\omega)\}\biggr \}.
\label{eq: l22kk_final}
\end{eqnarray}
The derivation for $L_{22}^{KP}=L_{22}^{PK}$ is identical to that of
$L_{12}^P$ except we have an extra factor of $(\epsilon-\mu)$ in
Eq.~(\ref{eq: l12p_real}).  Performing the integration over $\epsilon$
then produces
\begin{eqnarray}
L_{22}^{KP}&=&\frac{T\sigma_0}{e^2}\int_{-\infty}^{\infty}d\omega
\left ( -\frac{df(\omega)}{d\omega}\right )\cr
&\times&\biggr \{ [\omega-{\rm Re}\Sigma(\omega)]{\rm Re}\Sigma(\omega)
\tau(\omega)-{\rm Im}
G(\omega){\rm Im}\Sigma(\omega)\cr
&+&2[{\rm Im}\Sigma(\omega)]^2-2[{\rm Im}\Sigma(\omega)]^2{\rm Re} \{
[\omega+\mu-\Sigma(\omega)]G(\omega)\}\cr
&+&2[\omega-2{\rm Re}\Sigma(\omega)]{\rm Im}\Sigma(\omega){\rm Im}
\{[\omega+\mu-\Sigma(\omega)]G(\omega)\}\biggr \}.
\label{eq: l22kp_final}
\end{eqnarray}

The final term we must evaluate is $L_{22}^{PP}$.  This is the most complicated
term to evaluate and we are unable to do so following the same strategy as
employed in the $L_{12}^P$ derivation.   Instead, we proceed by an alternate
method based on the equation of motion (EOM) technique.  The EOM for the 
Fermionic creation and annihilation operators (in the momentum basis) are
\begin{equation}
\frac{\partial}{\partial \tau}c_{q\sigma}^\dagger(\tau)=[\epsilon(q)-\mu]
c_{q\sigma}^\dagger(\tau)+U\sum_kW(k)c_{q+k\sigma}^\dagger(\tau),
\label{eq: eomc+}
\end{equation}
and
\begin{equation}
\frac{\partial}{\partial \tau}c_{q\sigma}(\tau)=-[\epsilon(q)-\mu]
c_{q\sigma}(\tau)-U\sum_kW(k)c_{q-k\sigma}(\tau).
\label{eq: eomc}
\end{equation}  
These EOMs can be employed to express the correlation function of the
heat-current operators in terms of derivatives with respect to
imaginary time as shown below
\begin{eqnarray}
&~&\bar L_{22}^{PP}(i\nu_l)=\frac{\pi TU^2}{4}\int_0^\beta d\tau e^{i\nu_l\tau}
\sum_{qq^\prime q^{\prime\prime}q^{\prime\prime\prime}\sigma\sigma^\prime}\cr
&~&
({\bf v}_{q\alpha}+{\bf v}_{q^{\prime\prime\prime}\alpha})
({\bf v}_{q^\prime \beta}+{\bf v}_{q^{\prime\prime}\beta}) \cr
&\times& \langle T_\tau W(q-q^{\prime\prime\prime})W(q^\prime-q^{\prime\prime})
c^\dagger_{q\sigma}(\tau)c_{q^{\prime\prime\prime}\sigma}(\tau)\cr
&~&
c^{\dagger}_{q^\prime \sigma^\prime}(0)c_{q^{\prime\prime}\sigma^{\prime}}(0)
\rangle\cr
&=&\pi T\int_0^\beta d\tau e^{i\nu_l\tau}
\sum_{qq^\prime \sigma\sigma^\prime}{\bf v}_{q\alpha}{\bf v}_{q^\prime \beta}
\lim_{\tau^\prime\rightarrow\tau^-}\lim_{\tau^{\prime\prime\prime}
\rightarrow\tau^{\prime\prime -}\rightarrow 0^+}\cr
&~&\langle T_\tau [\{\frac{1}{2}(\partial_\tau-\partial_{\tau^\prime})-
(\epsilon_{q^\prime}-\mu)\} c^\dagger_{q^\prime\sigma}(\tau)
c_{q^\prime\sigma}(\tau^\prime)]\cr
&\times&[\{\frac{1}{2}(\partial_{\tau^{\prime\prime}}-
\partial_{\tau^{\prime\prime\prime}})-
(\epsilon_{q}-\mu)\} c^\dagger_{q\sigma^\prime}(\tau^{\prime\prime})
c_{q\sigma^\prime}(\tau^{\prime\prime\prime})]\rangle.
\label{eq: l22pp}
\end{eqnarray}
Now each of the operator averages can be expressed in terms of Green's 
functions, since the velocity factors guarantee there will be no vertex
corrections.  Noting further, that the integrals will only contribute
if $\alpha=\beta$ finally yields 
\begin{eqnarray}
&~&\bar L_{22}^{PP}(i\nu_l)=\pi T \int_0^\beta d\tau e^{i\nu_l\tau} 
\sum_{q\sigma}v_{q\alpha}^2\cr
&~&\times\biggr [ \frac{1}{2}\partial_\tau G_{q\sigma}(\tau)\partial_\tau 
G_{q\sigma}(-\tau)
-\frac{1}{4}\partial_\tau^2G_{q\sigma}(\tau)G_{q\sigma}(-\tau)\cr
&~&\quad-\frac{1}{4}G_{q\sigma}(\tau)\partial_\tau^2G_{q\sigma}(-\tau)\cr
&~&\quad+(\epsilon_q-\mu)\{ G_{q\sigma}(\tau)\partial_\tau G_{q\sigma}(-\tau)-
\partial_\tau G_{q\sigma}(\tau)G_{q\sigma}(-\tau)\}\cr
&~&\quad-(\epsilon_q-\mu)^2G_{q\sigma}(\tau)G_{q\sigma}(-\tau)\biggr ].
\label{eq: l22pp_tau_deriv}
\end{eqnarray}
We need to be able to produce expressions for the derivatives of the
Green's functions.  We do so by first writing the Green's function as a
Fourier series over the Matsubara frequencies, and then taking the
derivative into the Matsubara summation.  This may appear to be
mathematically unsound, but we do so by adding and subtracting the
known $\delta(\tau)$ behavior of the derivative, to regularize the
summation.  Since we are interested only in $0<\tau<\beta$, this procedure
has no convergence issues.  Likewise, one is also able to take the second
derivative in this fashion.  We find for $0<\tau <\beta$
\begin{eqnarray}
\partial_\tau G_{q\sigma}(\tau)&=&
-(\epsilon_q-\mu)G_{q\sigma}(\tau)\cr
&-&T\sum_me^{-i\omega_m\tau}\frac
{\Sigma_{m\sigma}}{i\omega_m+\mu-\Sigma_{m\sigma}-\epsilon_q},\cr
\partial_\tau^2 G_{q\sigma}(\tau)&=&
+(\epsilon_q-\mu)^2G_{q\sigma}(\tau)\cr
&+&(\epsilon_q-\mu)T\sum_me^{-i\omega_m\tau}\frac
{\Sigma_{m\sigma}}{i\omega_m+\mu-\Sigma_{m\sigma}-\epsilon_q}\cr
&+&T\sum_me^{-i\omega_m\tau}\frac
{i\omega_m\Sigma_{m\sigma}}{i\omega_m+\mu-\Sigma_{m\sigma}-\epsilon_q},
\label{eq: g_tau_deriv}
\end{eqnarray}
with similar formulae for $G_{q\sigma}(-\tau)$.
Substituting the derivatives from Eq.~(\ref{eq: g_tau_deriv}) into
Eq.~(\ref{eq: l22pp_tau_deriv}), and then simplifying the result finally
produces
\begin{eqnarray}
&~&\bar L_{22}^{PP}(i\nu_l)=-\frac{\pi T^2}{4}\sum_n\sum_{q\sigma}v_{q\alpha}^2
\cr
&~&\times\biggr [\{\Sigma_\sigma(i\omega_n)+\Sigma_\sigma(i\omega_{n+l})\}^2
G_{q\sigma}(i\omega_n)G_{q\sigma}(i\omega_{n+l})\cr
&~&+\Sigma_\sigma(i\omega_{n+l})G_{q\sigma}(i\omega_n)+
\Sigma_\sigma(i\omega_n)G_{q\sigma}(i\omega_{n+l})\biggr ].
\label{eq: l22pp_imag_final}
\end{eqnarray}
It is easy to understand the first terms in this expression, as they are
what one would naively recover when following the same Wick analysis that
was done previously for $L_{12}^P$.  We have not been able to discover a
direct operator-based
derivation of the second terms, but they are critical for providing the
right answer for $L_{22}^{PP}$.  Performing
the analytic continuation and simplifying yields our final result
\begin{eqnarray}
L_{22}^{PP}&=&\frac{T\sigma_0}{e^2}\int_{-\infty}^{\infty}d\omega
\left ( -\frac{df(\omega)}{d\omega}\right )\cr
&\times&\biggr \{ [{\rm Re}\Sigma(\omega)]^2\tau(\omega)-
2[{\rm Im}\Sigma (\omega)]^2+{\rm Im}\Sigma(\omega){\rm Im}G(\omega)\cr
&+&2[{\rm Im}\Sigma(\omega)]^2{\rm Re}\{[\omega+\mu-\Sigma(\omega)]G(\omega)\}
\cr
&+&4{\rm Re}\Sigma(\omega){\rm Im}\Sigma(\omega){\rm Im}\{ 
[\omega+\mu-\Sigma(\omega)]G(\omega)\}\biggr \}.
\label{eq: l22pp_final}
\end{eqnarray}
Summing together Eq.~(\ref{eq: l22kk_final}), twice Eq.~(\ref{eq: l22kp_final}),
and Eq.~(\ref{eq: l22pp_final}) gives the Mott form
\begin{equation}
L_{22}=\frac{T\sigma_0}{e^2}\int_{-\infty}^{\infty}d\omega
\left ( -\frac{df(\omega)}{d\omega}\right )
\tau(\omega)\omega^2.
\label{eq: l22_final}
\end{equation}

We can also
generalize the original Jonson-Mahan argument to prove relations between
$L_{21}$ and $L_{11}$ and between $L_{22}$ and $L_{12}$. Our method is
different from their proof and relies on the infinite-dimensional limit,
but one could proceed in their fashion if desired. We begin with the 
generalized two-particle correlation function
\begin{eqnarray}
&~&
F_{\alpha\beta}(\tau,\tau^\prime,\tau^{\prime\prime},\tau^{\prime\prime\prime})
=\cr
&~&\sum_{qq^\prime\sigma\sigma^\prime}{\bf v}_{q\alpha}{\bf v}_{q^\prime\beta}
\langle T_\tau c^\dagger_{q\sigma}(\tau)c_{q\sigma}(\tau^\prime)
c^\dagger_{q^\prime\sigma^\prime}(\tau^{\prime\prime})c_{q^\prime\sigma^\prime}
(\tau^{\prime\prime\prime})\rangle.
\end{eqnarray}
In the infinite-dimensional limit, the two-particle correlation function
is expressed by just its bare bubble because the irreducible charge vertex
has a different symmetry than ${\bf v}_q$.  Hence, we immediately learn
that
\begin{eqnarray}
&~&F_{\alpha\beta}(\tau,\tau^\prime,\tau^{\prime\prime},\tau^{\prime\prime\prime})\cr
& =&-\sum_{q\sigma}{\bf v}_{q\alpha}^2\delta_{\alpha\beta}
G_{q\sigma}(\tau^{\prime\prime\prime}-\tau)
G_{q\sigma}(\tau^{\prime}-\tau^{\prime\prime}).
\label{eq: F_wick}
\end{eqnarray}
But
\begin{equation}
G_{q\sigma}(\tau)=\int d\omega A(k,\omega)e^{-\omega\tau}[1-f(\omega)],
\end{equation}
for $\tau >0$ and
\begin{equation}
G_{q\sigma}(\tau)=\int d\omega A(k,\omega)e^{-\omega\tau}[-f(\omega)],
\end{equation}
for $\tau <0$. Substituting into Eq.~(\ref{eq: F_wick}), then yields
\begin{eqnarray}
&~&F_{\alpha\beta}(\tau,\tau^\prime,\tau^{\prime\prime},\tau^{\prime\prime\prime})
=\frac{\delta_{\alpha\beta}}{2d}\int d\epsilon \rho(\epsilon)\int d\omega
\int d\omega^\prime\cr
&\times&A(\epsilon,\omega)A(\epsilon,\omega^\prime)
e^{\omega(\tau-\tau^{\prime\prime\prime})-\omega^\prime(\tau^\prime-
\tau^{\prime\prime})}f(\omega)[1-f(\omega^\prime)].
\label{eq: f_ac}
\end{eqnarray}
Using this function we can construct the relevant ``polarization operators''.
Recalling the EOM in Eqs.~(\ref{eq: eomc+}) and (\ref{eq: eomc})
shows that
\begin{equation}
\lim_{\tau^\prime\rightarrow\tau^-}\frac{1}{2}\left (
\frac{\partial}{\partial\tau}-\frac{\partial}{\partial\tau^\prime}\right )
\sum_{q\sigma}{\bf v}_q
c^\dagger_{q\sigma}(\tau)c_{q\sigma}(\tau^\prime)={\bf j}_Q(\tau).
\label{eq: heat_current_identity}
\end{equation}
The Jonson-Mahan theorem will hold for any Hamiltonian that satisfies
Eq.~(\ref{eq: heat_current_identity}) (which follows from 
the relevant commutators and equations of
motion). This identity holds true for the Falicov-Kimball
model, the static Holstein model, and the periodic Anderson model\cite{mahan},
but it does not hold for either the Hubbard model or the quantum Holstein
model, a fact which does not appear to be widely known.
The ``polarization operators'' then become
\begin{equation}
\bar L_{11}=\pi T\int_0^\beta e^{i\nu_l\tau}F(\tau,\tau^-,0,0),
\end{equation}
for the conductivity,
\begin{equation}
\bar L_{12}=\pi T\int_0^\beta e^{i\nu_l\tau}\frac{1}{2}\left ( \frac{\partial}
{\partial\tau}-\frac{\partial}{\partial\tau^\prime}\right )
F(\tau,\tau^\prime,0,0),
\end{equation}
(in the limit where $\tau^\prime\rightarrow \tau^-$) for the thermopower, and
\begin{eqnarray}
\bar L_{22}&=&\pi T\int_0^\beta e^{i\nu_l\tau}\frac{1}{4}\left ( \frac{\partial}
{\partial\tau}-\frac{\partial}{\partial\tau^\prime}\right )
\left ( \frac{\partial}
{\partial\tau^{\prime\prime}}-\frac{\partial}{\partial\tau^{\prime\prime\prime}}\right )\cr
&\times&F(\tau,\tau^\prime,\tau^{\prime\prime},\tau^{\prime\prime\prime}),
\end{eqnarray}
(in the limit where $\tau^\prime\rightarrow \tau^-$,
$\tau^{\prime\prime\prime}\rightarrow \tau^{\prime\prime -}$, and
$\tau^{\prime\prime}\rightarrow 0^+$) for the thermal conductivity. Because of
Eq.~(\ref{eq: f_ac}),
the analytic continuation is trivial (one first converts from imaginary time to
Matsubara frequencies and then performs the Wick rotation to the real frequency
axis), and if we note the identity
\begin{equation}
f(\omega)-f(\omega+\nu)=-f(\omega)[1-f(\omega+\nu)][e^{-\beta\nu}-1],
\end{equation}
then we can easily compute that
\begin{equation}
L_{11}=\frac{T\sigma_0}{e^2}\int d\epsilon\rho(\epsilon)\int d\omega
\left (-\frac{df(\omega)}{d\omega}\right )A^2(\epsilon,\omega),
\end{equation}
for the conductivity,
\begin{equation}
L_{12}=\frac{T\sigma_0}{e^2}\int d\epsilon\rho(\epsilon)\int d\omega
\left (-\frac{df(\omega)}{d\omega}\right )A^2(\epsilon,\omega)\omega,
\end{equation}
for the thermopower,
and 
\begin{equation}
L_{22}=\frac{T\sigma_0}{e^2}\int d\epsilon\rho(\epsilon)\int d\omega
\left (-\frac{df(\omega)}{d\omega}\right )A^2(\epsilon,\omega)\omega^2,
\end{equation}
for the thermal conductivity.  This proves Mott's form for the thermal
transport [since 
$\int d\epsilon \rho(\epsilon) A^2(\epsilon,\omega)=\tau(\omega)$].

\section{Numerical Results}

We present results first for a case where the filling of the localized
electrons is a constant.  We choose the symmetric case of $\langle w\rangle=1/2$
and half filling $\rho_e=1$ for the electrons.  We perform calculations for 
two cases: (i) $U=1$ which is a moderately correlated metal and (ii) $U=2$ which
is a strongly correlated insulator.

\begin{figure}[htbf]
\epsfxsize=3.0in
\centerline{\epsffile{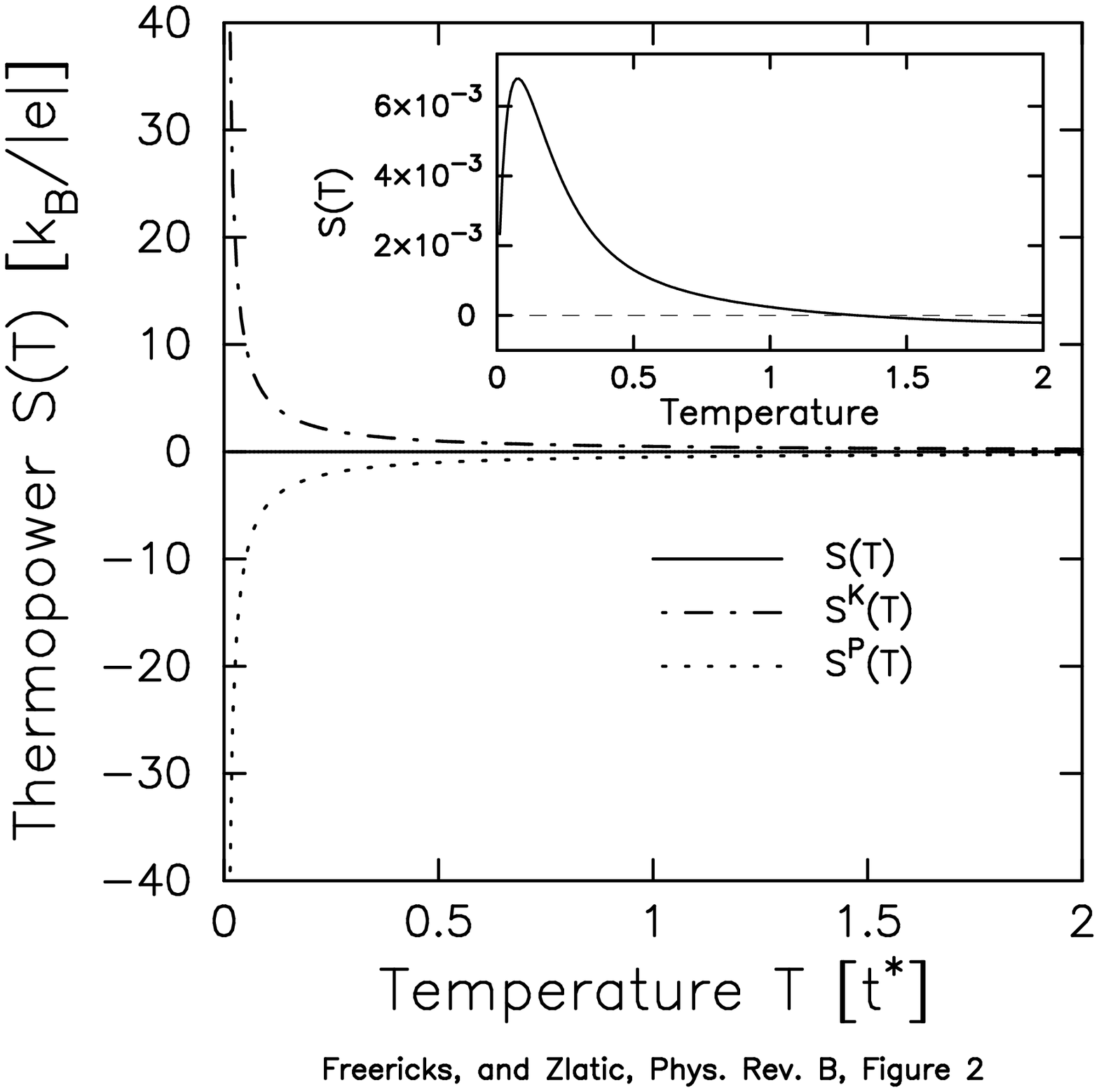}}
\caption{Thermopower for the case $U=1$, $w_1=0.5$, and $\rho_d=1$. The
main panel shows the two different contributions from the kinetic and
potential energy pieces of the heat
current and the inset shows the net thermopower.  Note how the two
pieces are large and nearly cancel to produce $S$, and how there is a
sign change near $T\approx 1.4$.
\label{fig: tp_u=1}}
\end{figure}

\begin{figure}[htbf]
\epsfxsize=3.0in
\centerline{\epsffile{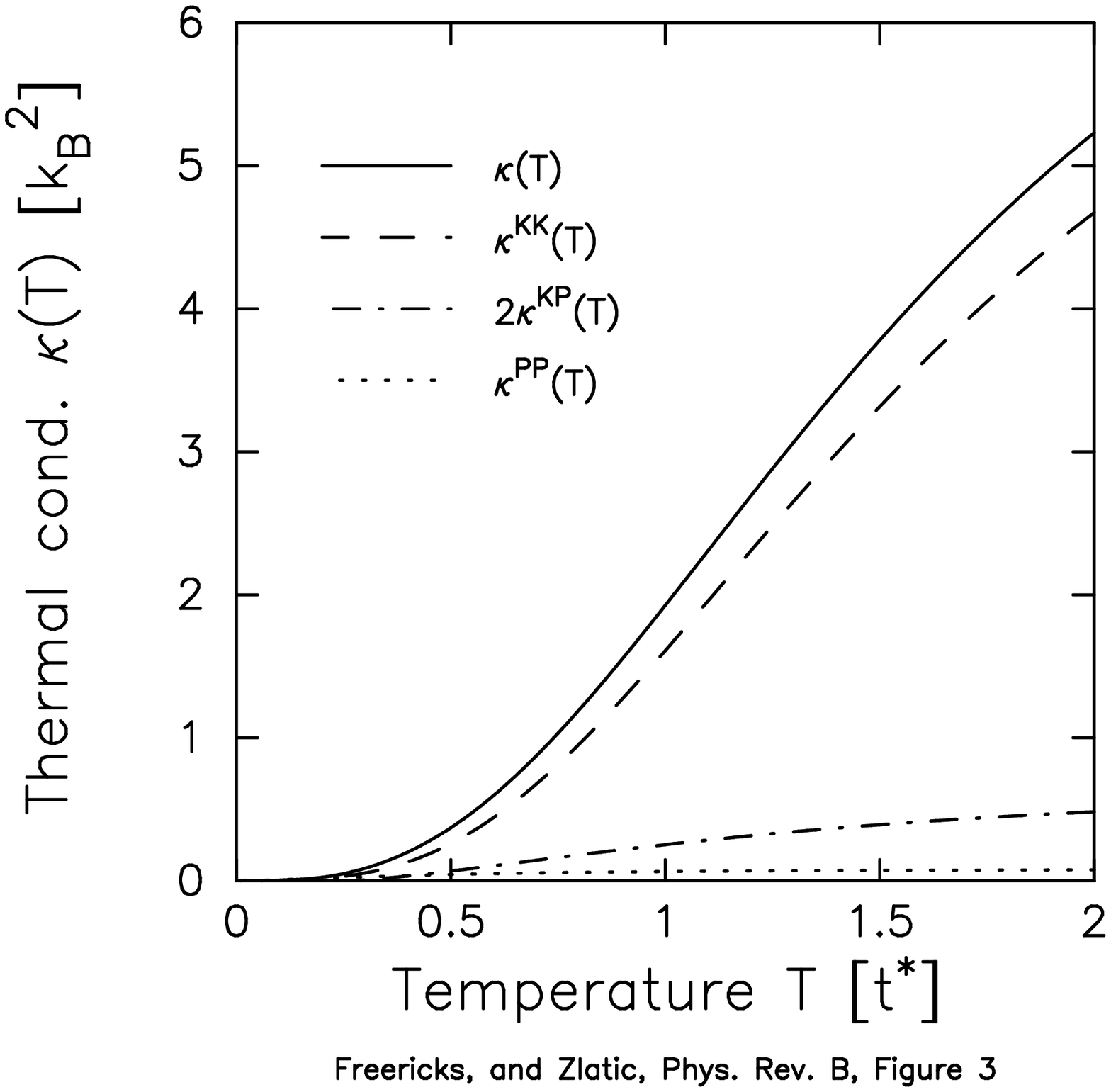}}
\caption{Thermal conductivity for the case $U=1$, $w_1=0.5$, and $\rho_d=1$. The
plot shows the different contributions from the kinetic and
potential energy pieces of the heat current. Note how the thermal 
conductivity is essentially described
by the kinetic-energy-only piece for moderate correlation strength.
\label{fig: tc_u=1}}
\end{figure}

The thermopower behaves as expected---it vanishes for large and small 
temperature, it has an electron-like peak at lower temperatures and a sign
change at a temperature on the order of half the bandwidth. What is 
surprising is that there is such a
large compensation between the kinetic and potential energy pieces of
the thermopower to produce the net thermopower (note the three order of 
magnitude difference in the
scales for the main figure and the inset).
The thermal conductivity also appears as expected. We can
see that while the contributions from the potential energy are critical in
determining the right thermopower, they have a relatively mild effect in
the thermal conductivity for a moderately correlated metal (note how close
the total thermal conductivity is to the kinetic-energy-only contribution).

As we increase the correlation strength, so that the interacting density of
states has a gap and the system is a correlated insulator, the
behavior of the thermal transport changes.  

\begin{figure}[htbf]
\epsfxsize=3.0in
\centerline{\epsffile{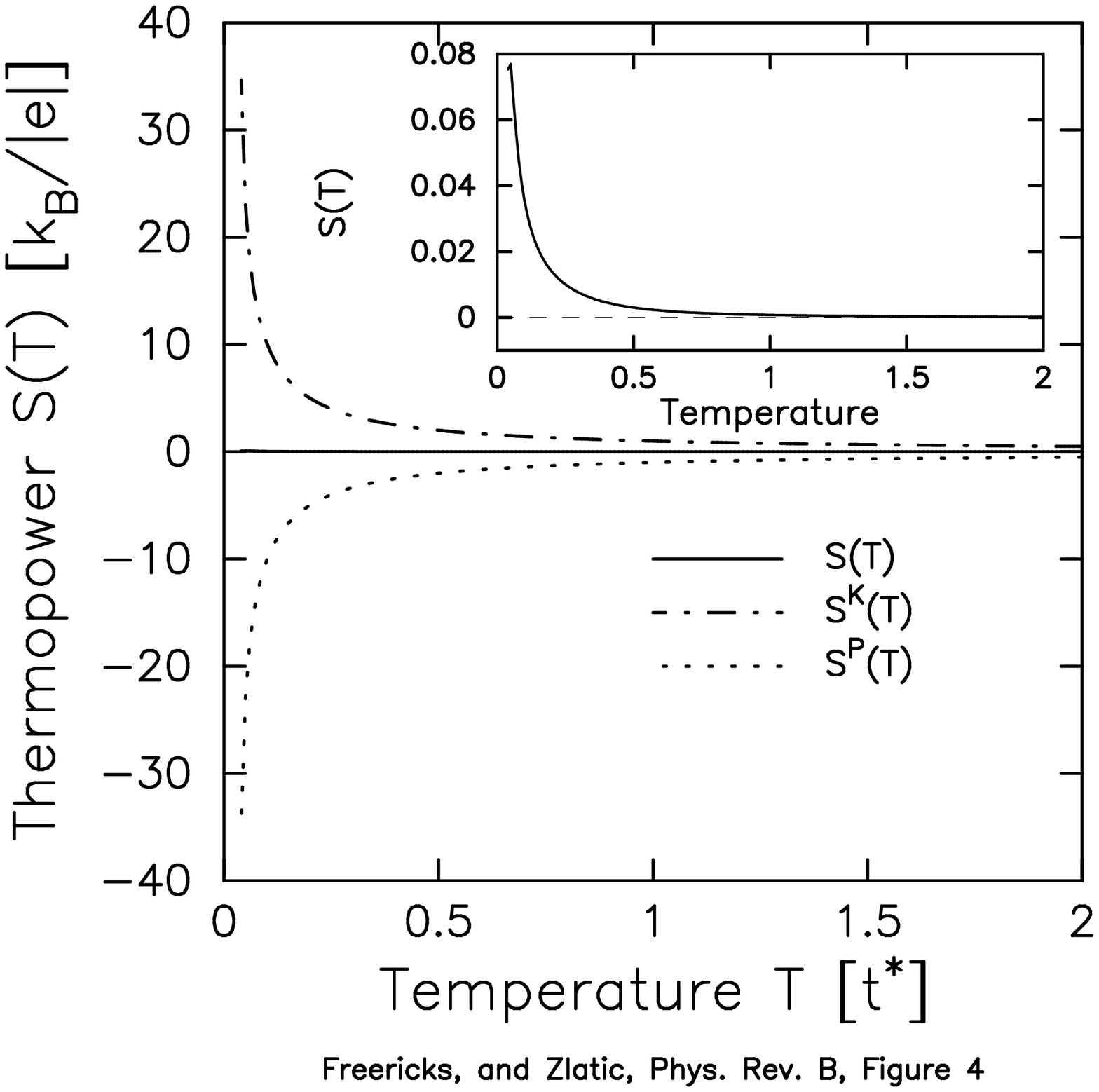}}
\caption{Thermopower for the case $U=2$, $w_1=0.5$, and $\rho_d=1$. The
main panel shows the two different contributions from the kinetic and
potential energy pieces of the heat current and the inset shows the 
net thermopower.  Note how the two
pieces are large and nearly cancel to produce $S$ and how the thermopower
appears to diverge as $T$ becomes small (calculations run into numerical
problems as $T\rightarrow 0$).
\label{fig: tp_u=2}}
\end{figure}

\begin{figure}[htbf]
\epsfxsize=3.0in
\centerline{\epsffile{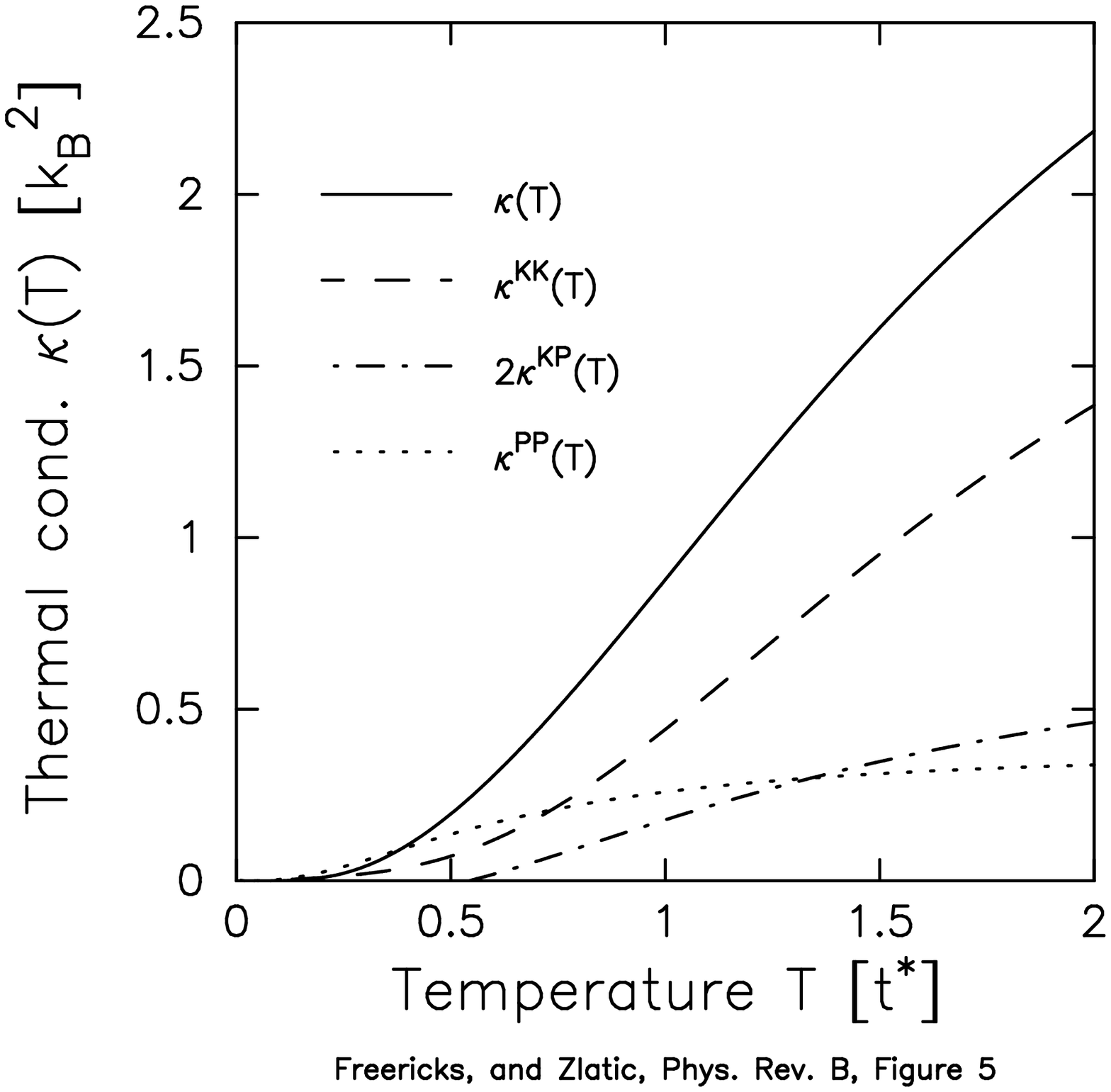}}
\caption{Thermal conductivity for the case $U=2$, $w_1=0.5$, and $\rho_d=1$. The
plot shows the different contributions from the kinetic and
potential energy pieces of the heat current. Note how the potential energy 
terms become increasingly important.
\label{fig: tc_u=2}}
\end{figure}

The thermopower has the characteristic insulating behavior here, with
what appears to be a divergence as $T\rightarrow 0$. The divergence arises from 
the presence of a gap in the single-particle spectrum---both $L_{11}$ and 
$L_{12}$ approach zero exponentially in $T$ (with the same exponent), but
the ratio is proportional to the size of the gap, so $S\rightarrow C/T +B$ as 
$T\rightarrow 0$ (similar results have been found when applying scaling theory
to the Anderson transition\cite{villagonzalo}).  Here
this is ``allowed'' thermodynamically, because the ``ground-state''
of the insulating phase has nonzero entropy, since we forced the system
into the paramagnetic insulating phase.  In a real system, however, there must
be a transition to a ground state where the entropy is quenched.  In such
a case, one would expect to see a large peak in the thermopower at low
energies, with the thermopower 
ultimately going to zero as $T\rightarrow 0$.  We know
of no real correlated insulator that has a diverging thermopower as 
$T\rightarrow 0$.

The computation
at very small values of $T$ is difficult because the electrical conductivity
(or equivalently $L_{11}$) approaches zero and there are numerical difficulties
associated with properly calculating the conductivity in this regime due
to the cancellation of two large and nearly equal numbers. Since the thermopower
requires the electrical conductivity, if that cannot be calculated
accurately then spurious behavior will be seen in the thermopower. The thermal
conductivity looks similar to the weaker correlated case, except it appears
to go to zero at a nonzero temperature which is the expected behavior for
a correlated insulator with a gap and the potential energy pieces become
increasingly more important (particularly at low temperature).  

In both of these half-filled cases, that of a correlated metal and a correlated
insulator, the low-temperature
thermopower is determined by a slightly larger contribution
from the kinetic energy piece of the heat current than from the potential
energy piece of the heat current.  The thermal conductivity, on the other
hand has an evolution of going from a result nearly completely determined
by the kinetic energy only piece of the heat current correlation functions
to one where the potential energy pieces of the heat current contribute
progressively more and more to the total thermal conductivity.

Next we present results for the case where the total filling
$\rho_e+\langle
w\rangle=1.5$ is a constant but the electrons can change from localized
to itinerant (i.e., we fix the total electron concentration not the
individual electron concentrations). We choose values of the
parameters\cite{zlatic_bled} where
the system has a sharp transition from a state at high temperature that
has large f-occupancy ($\langle w\rangle\approx 0.36$ for
$0.2<T<0.8$), to a state at low
temperature with no f-electrons (the crossover occurs near $T=0.04$).
We find that the results do not depend too strongly on the parameters in
this regime, and choose $E_f=-0.7$ and $U=4$ as a canonical system that
is similar to YbInCu$_4$. The main difference from the symmetric case
studied above is that the localized electron filling goes to zero as
$T\rightarrow 0$.  Hence both the kinetic and potential contributions to
the thermopower become small in this limit, and there is no large
cancellation between two nearly equal numbers to determine the
thermopower.
We see that in the thermal conductivity, the contributions from the mixed
kinetic and potential energy pieces are the most important, which is
an indication of the strengthening of the correlations in the system. This
feature is hard to see from the shape of the thermal conductivity itself.

\begin{figure}[htbf]
\epsfxsize=3.0in
\centerline{\epsffile{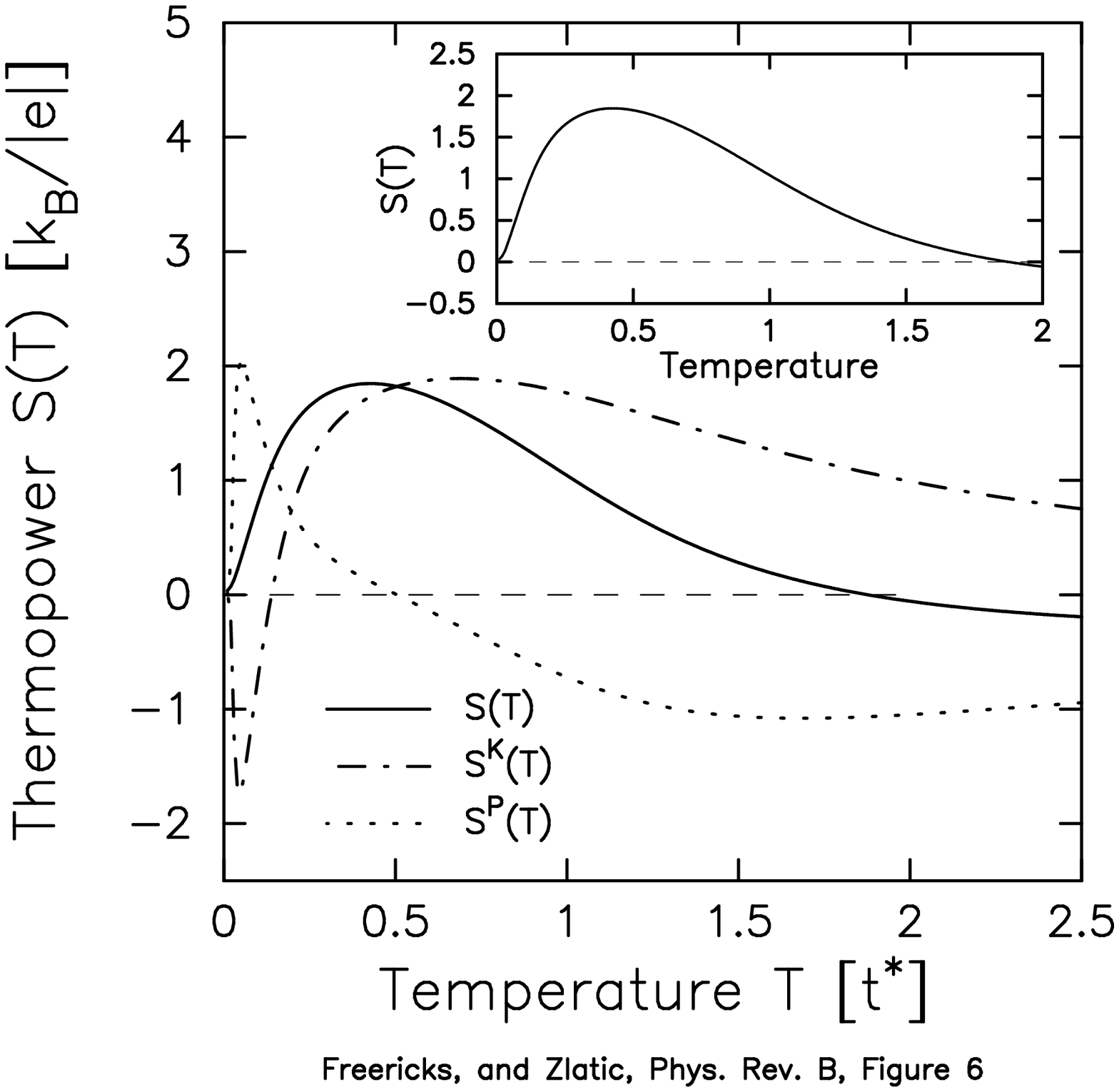}}
\caption{Thermopower for the case $U=4$, $w_1+\rho_d=1.5$, and
$E_F=-0.7$. The
main panel shows the two different contributions from the kinetic and 
potential energy pieces of the heat current and the inset shows the net
thermopower.  Note the absolute scale for the thermopower is much larger
here.
\label{fig: tp_ef=-0.7}}
\end{figure}

\begin{figure}[htbf]
\epsfxsize=3.0in
\centerline{\epsffile{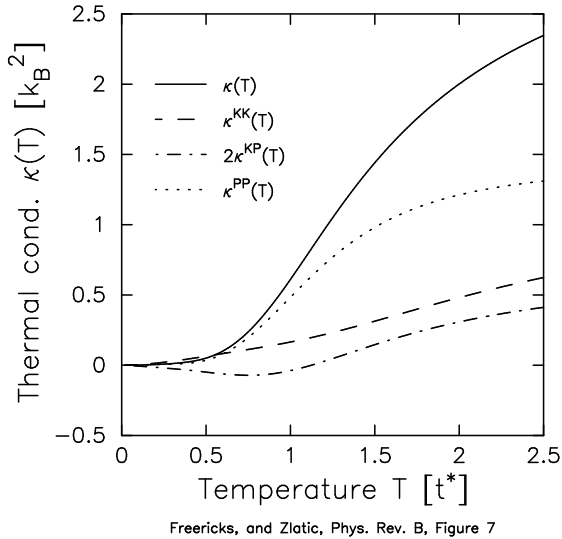}}
\caption{Thermal conductivity for the case $U=4$, $w_1+\rho_d=1.5$, and
$E_F=-0.7$. The
plot shows the different contributions from the kinetic and
potential energy pieces of the heat current.
\label{fig: tc_ef=-0.7}}
\end{figure}

The thermopower has an interesting exchange of importance of the
different pieces of the heat current as temperature is varied.  For high
temperatures, the thermopower is determined by both parts of the heat
current
and a compensation effect is important.  In the moderate temperature
regime,
the thermopower is dominated by kinetic-energy pieces, which then give way
to the potential energy domination at low temperature, that eventually
shrinks as $T\rightarrow 0$ and the thermopower vanishes. 

Note, would need to reverse the sign of the thermopower to describe
YbInCu$_4$, since its charge carriers are holes rather than electrons.  
We should also remark that the thermopower of the Falicov-Kimball model 
does not have a low-energy peak associated with the ``valence-change''
transition---such a sharp peak occurs in YbInCu$_4$-like systems 
because of hybridization effects not included in this model.
Consequently, the thermopower also does not have a low-temperature
sign change as seen in the experimental
data\cite{yic_thermopower} (a sign change occurs at $T\approx 2$),
but we see that if one could reduce the potential energy piece of the  
thermal current, then the kinetic energy contributions to the thermopower
could
cause a sign change to occur.  This cannot happen in a pure
Falicov-Kimball
model though, because of the Jonson-Mahan theorem.  The sign change for the
pure Falicov-Kimball model generically
occurs at much larger values of temperature, on the order of the
bandwidth.
As regards the
thermal conductivity, we find $\kappa$ is dominated by the
potential-energy
piece as $T\rightarrow 0$ and the mixed piece yields a negative
contribution
over a wide temperature range.

\section{Conclusions}

We have examined thermal transport in the spin-one-half Falicov-Kimball model.
We chose this model because the transport properties can be solved exactly,
and they provide an alternate proof of the Jonson-Mahan theorem for the
thermopower.  We provide the proof in two different ways.  The first is
a brute-force application of the dynamical mean field theory to calculate
all relevant correlation functions and combine all terms to yield the
final expressions for the thermal transport coefficients.  The second is
based largely on the techniques of Jonson and Mahan, but one can determine
the important ``generalized polarization'' functions exactly in the large
dimensional limit.  Here we extend the Jonson-Mahan arguments to show 
analogous results hold for the thermal conductivity as well.

Our formulation also allows us to decompose the contributions
to the thermopower and the thermal conductivity into the respective
contributions from the kinetic energy piece and the potential energy piece
of the thermal current.  We find that generically, these pieces are large
and opposite in sign for the thermopower so that thermal transport carried
by the kinetic heat current is almost completely compensated by the potential
heat current producing a small net thermopower.  For the thermal conductivity,
we see an evolution of the transport being dominated first by kinetic energy
terms and then potential energy terms as the strength of the correlations 
increase.  We note, that because the kinetic-energy contribution to the
thermopower can be straightforwardly determined for a number of models,
any Hamiltonian that satisfies the Jonson-Mahan theorem can be separated
into its kinetic and potential pieces for the thermopower by simply subtracting
the kinetic energy piece from the Jonson-Mahan result.

\acknowledgments
We would like to acknowledge stimulating discussions with G. Czycholl,
B. Letfulov, G. Mahan, R. R\"omer and C. Villagonzalo.
This work was supported by the National Science Foundation
under grant DMR-9973225.


\end{document}